\documentclass[pre,a4paper,oneside,twocolumn]{revtex4}

\usepackage{graphicx}
\usepackage[brazil]{babel}
\usepackage[latin1]{inputenc}
\usepackage{lastpage}

\usepackage{graphicx}  
\usepackage{latexsym}   
\usepackage{color}
\begin{document}

\title{The influence of memory in deterministic walks in random media: analytical calculation within a mean field approximation}

\author{\firstname{C\'esar} Augusto Sangaletti \surname{Ter\c{c}ariol}} 
\email{cesartercariol@gmail.com}
\affiliation{Faculdade de Filosofia, Ci\^encias e Letras de Ribeir\~ao Preto, \\
             Universidade de S\~ao Paulo \\ 
             Avenida Bandeirantes, 3900 \\ 
             14040-901, Ribeir\~ao Preto, SP, Brazil}

\affiliation{Centro Universit\'ario Bar\~ao de Mau\'a \\
             Rua Ramos de Azevedo, 423 \\ 
             14090-180, Ribeir\~ao Preto, SP, Brazil}

\author{\firstname{Alexandre} Souto \surname{Martinez}}
\email{asmartinez@ffclrp.usp.br}

\affiliation{Faculdade de Filosofia, Ci\^encias e Letras de Ribeir\~ao Preto, \\
             Universidade de S\~ao Paulo \\ 
             Avenida Bandeirantes, 3900 \\ 
             14040-901, Ribeir\~ao Preto, SP, Brazil}
\date{\today}

\begin{abstract}

Consider a random medium consisting of points randomly distributed so that there is no correlation among the distances. 
This is the random link model, which is the high dimensionality limit (mean field approximation) for the euclidean random point structure. 
In the random link model, at discrete time steps, the walker moves to the nearest site, which has not been visited in the last $\mu$ steps (memory), producing a deterministic partially self avoiding walk (the tourist walk).
We have obtained analitically the distribution of the number $n$ of points explored by a walker with memory $\mu = 2$, as well as the transient and period joint distribution.
This result enables to explain the abrupt change in the exploratory behavior between the cases $\mu = 1$ (memoryless, driven by extremal statistics) and $\mu = 2$ (with memory, driven by combinatorial statistics). 
In the $\mu = 1$ case, the mean newly visited points in the thermodynamic limit $(N \gg 1)$ is just $\langle n \rangle = e = 2.72 \ldots$ while in the $\mu = 2$ case, the mean number $\langle n \rangle$ of visited points is proportional to $N^{1/2}$.
Also, this result allows us to stabilish an equivalence between the random link model with $\mu=2$ and random map (uncorrelated back and forth distances) with $\mu=0$ and the drastic change between the cases where the transient time is null compared to non-null transient times.
\end{abstract}



\maketitle


\section{Introduction}

Although not as thoroughly studied as random walks in disordered media~\cite{fisher:1984} and complex media~\cite{metzler_2000}, which constitute an interesting problem for Physics, deterministic walks in regular~\cite{grassberger:92,gale:95} and disordered media~\cite{bunimovich:2004,lam_2006,santos:061114} present very interesting results, as an application to foraging~\cite{boyer_2004,boyer_2005,boyer_2006}.
The memory in random walks has the effect of changing the behavior of the gaussian displacement distribution~\cite{cressoni:070603}.  
Here, we are interested in understanding fundamental aspects of a partially self-avoiding deterministic walk algorithm, known as the tourist walk (TW)~\cite{lima_prl2001,stanley_2001,kinouchi:1:2002}. 
These walks, that are described below, have been applied to characterize thesaurus~\cite{kinouchi:1:2002}, as a pattern recognition algorithm~\cite{campiteli_2006} and image analysis~\cite{bruno_2006,backes_2006}. 

Consider $N$ points (sites, cities) randomly distributed inside a $d$-dimensional hypercube with unitary edges.
The distance $D_{i, j}$ between any two points $s_i$ and $s_j$ is calculated via euclidean metrics. 
The walker leaves a given point and moves obeying the deterministic rule of going to the nearest point (shortest euclidean distance), which has not been visited in the $\mu$ preceding steps.
This rule produces trajectories with an initial transient part of $t$ steps and a cycle of $p$ steps as a final periodic part.  
Once trapped in a cycle, the walker does not visit new points anylonger. 
Short transient times and short period cycles limit exploration of the medium by the walker. 

Analytical results~\cite{tercariol_2007_physa} could be obtained for (i) memoryless walkers in the deterministic~\cite{tercariol_2005} and stochastic~\cite{risaugusman:1:2003,martinez:1:2004} versions of the TW and for (ii) deterministic walk with arbitrary memory in one-dimensional systems~\cite{tercariol_2007}. 
Here we consider the memory effect in deterministic walks in a mean field approximation. 

The deterministic TW, with memory $\mu = 0$, is trivial since the walker does not move at each time step, so that the transient-time/period joint distribution is simply: $S^{(N)}_{0, d}(t, p) = \delta_{t, 0} \delta_{p, 1}$, where $\delta_{i, j}$ is the Kronecker delta. 
With memory $\mu = 1$, the walker must leave the current site at each time step. 
The joint distribution $S^{(N)}_{1, d}(t, p)$ is obtained considering the trajectories of a tourist leaving from all sites of a given map and statistics is performed for different realizations (maps). 
For $N \gg 1$, the transient-time/period joint distribution is obtained analytically for arbitrary dimensionality~\cite{tercariol_2005}: $S^{(\infty)}_{1, d}(t, p) = [(t + I_d^{-1}) \Gamma(1 + I_d^{-1})/\Gamma(t + p + I_d^{-1})] \delta_{p, 2}$, where $\Gamma(z)$ is the gamma function and $I_d = I_{1/4}[1/2,(d+1)/2]$ is the normalized incomplete beta function. 
This case does not lead to exploration of the random medium since after a short transient, the tourist gets trapped in pairs of cities that are mutually nearest neighbors. 

Interesting phenomena occur when the memory values are greater or equal to two ($\mu \ge 2$). 
In this case, the cycle distribution is no longer peaked at $p_{min} = \mu + 1$, but presents a whole spectrum of cycles with period $p \ge p_{min}$, with possible power-law decay~\cite{lima_prl2001,kinouchi:1:2002}, which favors exploration of the medium by the walker. 
The elucidation of this intringuing broadening of the cycle period distribution is our main objective. 

As the medium dimensionality $d$ incresases, the correlations between the distances $D_{i, j}$ become weaker and weaker, so that, in the high dimensionality limit ($d \rightarrow \infty$), the distances can be considered independent random variables, uniformly distributed in the interval $[0, 1]$~\cite{mezard:1986,percus:1996,percus:1997,percus:1998,percus:1999}.
This is the mean field model named Random Link (RL), where two euclidean constraints still remain: (i) the distance from a point to itself is null, $D_{i, i} = 0$, and (ii) the forward and backward distances are equal, $D_{i, j} = D_{j, i}$.
Breaking these constraints leads to the Random Map model (RM)~\cite{derrida:2:1997}, which is a mean field approximation for the Kauffman's model~\cite{kauffman:1969}. 
The neighborhood statistics for these models have being analytically studied in Ref.~\onlinecite{tercariol_2007_jpa}. 


In this paper, we obtain analytical results for the TW, with memory $\mu = 2$ in the $d \to \infty$ medium, i.e. the RL approximation. 
These results enable us to explain the main mechanism which makes the $\mu = 1$ and $\mu \ge 2$ situations so distinct. 
Also, they permit us to estabilish a relationship between the mean fields RL and RM models. 
The walks with memory $\mu=2$ in the symmetric independent random distance case (RL model) is equivalent to memoryless ($\mu=0$) walks in the assymmetric independent random distance case (RM model), which has been already solved in Ref.~\onlinecite{tercariol_2005}. 
Throughout this relationship between RL and RM models, we show that the decay for the cycle period distribution in the RL model is a power law $\propto p^{-1}$. 

Also we are able to explain the reason of the already observed numerically abrupt change in the in the transient/period joint distribution for null transient $t = 0$. 

The presentation of these results are briefly skechted in the following. 
In Sec.~\ref{Sec:DistrTildeN}, we calculate the probability $\tilde{S}_{2, rl}^{(N)}(\tilde{n})$ for the tourist, with memory $\mu=2$, to visit $\tilde{n}$ distinct sites before the first passage to any already visited site, walking on the RL model with $N$ sites. 
We start calculating the complementary cumulative distribution $\tilde{F}_{2, rl}^{(N)}(\tilde{n})$ (upper-tail distribution).
Next, throughout an analogy to the geometric distribution, we obtain the revisit $\tilde{p}_{2, rl}^{(N)}(j)$ (first passage) and exploration $\tilde{q}_{2, rl}^{(N)}(j)$ probabilities.
Using an alternative derivation, we obtain simpler expressions for these probabilities, which lead to a closed analytical expression for $\tilde{F}_{2, rl}^{(N)}(\tilde{n})$.
In Sec.~\ref{Sec:DistrN}, we show that the probability for the tourist to be trapped into a cycle when revisiting a site is 2/3, which is counterintuitive. 
This result (combined to previous ones) allows us to obtain the complementary cumulative distribution $F_{2, rl}^{(N)}(n)$ for the total number $n$ of visited sites (until the walker enters an attractor).
In Sec.~\ref{Sec:JointDistr}, we obtain the joint distribution $S_{2, rl}^{(N)}(t, p)$ of transient time $t$ and cycle period $p$ and show the drastic difference between the $t=0$ and $t \ne 0$ cases.
Final remarks are presented and future studies are proposed in Sec.~\ref{Sec:Conclusion}. 

\section{Distribution for the number of explored sites before the first passage (revisit)}
\label{Sec:DistrTildeN}

Consider that the tourist, who performs a walk with memory $\mu=2$ on the RL model with $N$ points, has visited $\tilde{n} \ge 3 = \mu+1 = \tilde{n}_{min}$ distinct sites and then revisits one of these sites.
Aiming to obtain the distribution $\tilde{S}_{2, rl}^{(N)}(\tilde{n})$ of the number $\tilde{n}$ of sites visited before the first passage, we start calculating the complementary cumulative (upper-tail) distribution
\begin{eqnarray*}
\tilde{F}_{2, rl}^{(N)}(\tilde{n}) = \sum_{k=\tilde{n}}^N \tilde{S}_{2, rl}^{(N)}(k)
\end{eqnarray*}
i.e., the probability for the tourist to explore at least $\tilde{n}$ distinct sites, before the first revisit.

\begin{figure}[htb]
\begin{center}
\includegraphics[angle=-90,scale = .5]{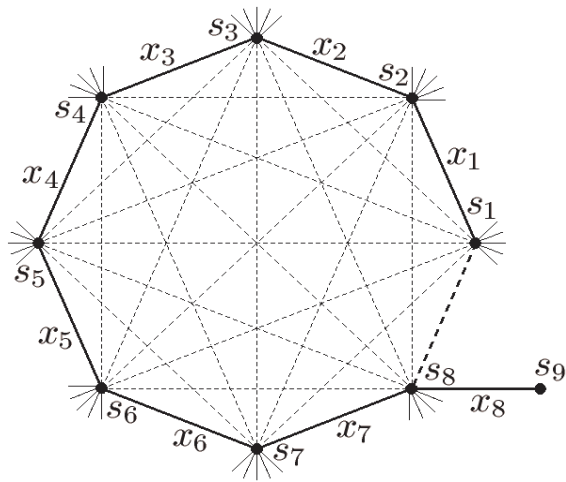}
\caption{Schematic representation of a walk with at least $\tilde{n}=9$ sites visited before the first passage.
The walker leaves from the site $s_1$ and follows the trajectory $s_1$, $s_2$, $s_3$, \ldots, $s_9$.}
\label{Fig:Trajectory}
\end{center}
\end{figure}

In the schema of Fig.~\ref{Fig:Trajectory}, the tourist leaves from a given site $s_1$ (first step, $j=1$) and follows the trajectory $s_1$, $s_2$, \ldots, $s_{\tilde{n}}$, exploring $\tilde{n}=9$ distinct sites, with no revisit.
For $1 \le i \le \tilde{n}-1$, let us denote
\begin{itemize}
\item $x_i$ the distance between the consecutive sites $s_i$ and $s_{i+1}$ in the trajectory (thick continuous lines of Fig.~\ref{Fig:Trajectory}),
\item $y_{i, k}$ the distances between the site $s_i$ in the trajectory and other sites outside the trajectory (thin continuous lines of Fig.~\ref{Fig:Trajectory}),
\item $z_{i, k}$ the distance between the non-consecutive sites $s_i$ and $s_k$ in the trajectory (slashed lines of Fig.~\ref{Fig:Trajectory}),
\end{itemize}
By definition of the RL model, all these distances $x_i$, $y_{i, k}$ and $z_{i, k}$ has uniform deviate in the interval $[0, 1]$.

The conditions for the tourist to follow the trajectory $s_1$, $s_2$, \ldots, $s_{\tilde{n}}$ in the first $\tilde{n}$ steps are
\begin{enumerate}
\item in the case $\mu=1$ (already solved in Ref.~\onlinecite{tercariol_2005}), the distances $x_i$ must obey the relation $x_{\tilde{n}-1} < x_{\tilde{n}-2} < \cdots < x_1$, once the tourist stops exploring new sites when $x_{i+1} > x_i$, giving rise to a cycle of period $p=2$.
But for the case $\mu=2$ addressed here, each distance $x_i$ may vary unrestrictly in the interval $[0, 1]$, because the memory $\mu=2$ forbids the tourist to move backward from $s_{i+1}$ to $s_i$ (even if $x_{i+1}>x_i$).
\item when the tourist is about to walk the distance $x_i$ (and move from $s_i$ to $s_{i+1}$) there exist $N-i$ non-explored sites at his/her disposal.
\item for each site $s_i$, all $N-(\tilde{n}-1)$ distances $y_{i, k}$ must be greater than $x_i$.
The probability for this to occur is $\left[ \int_{x_i}^1 \mbox{d}y_{i, k} \right]^{N-\tilde{n}+1} = (1-x_i)^{N-\tilde{n}+1}$.
The only exception is the site $s_{\tilde{n}-1}$, which has $N-\tilde{n}$ distances $y_{\tilde{n}-1, k}$ connected to it (see Fig.~\ref{Fig:Trajectory}, where $s_{\tilde{n}-1}$ corresponds to $s_8$).
\item to avoid shortcuts and revisits, each distance $z_{i, k}$ must be greater than both $x_i$ and $x_k$.
\end{enumerate}

These conditions lead to the following chained integrals:
\begin{eqnarray}
\tilde{F}_{2, rl}^{(N)}(\tilde{n}) & = & \prod_{i=1}^{\tilde{n}-2} \int_0^1 \mbox{d}x_i(N-i)(1-x_i)^{N-\tilde{n}+1} \nonumber \\ \nonumber \\
& & \int_0^1 \mbox{d}x_{\tilde{n}-1}(N-\tilde{n}+1)(1-x_{\tilde{n}-1})^{N-\tilde{n}} \nonumber \\ \nonumber \\
& & \prod_{i=1}^{\tilde{n}-3} \prod_{k=i+2}^{\tilde{n}-1} \int_{\mbox{max}(x_i, x_k)}^1 \mbox{d}z_{i, k} \; .
\label{Eq:ChainedIntegrals}
\end{eqnarray}
It is worthwhile to mention that we have made no approximation yet, hence Eq.~\ref{Eq:ChainedIntegrals} yields exact results even for small values of $N$, as Tab.~\ref{Tab:DistrN6} shows.

{\Large
\begin{table}[htb]
\begin{center}
\begin{tabular}{c|cc|cc|cc}
\hline
            &                                      &                                      &      & standard &            & dif. (in \\
$\tilde{n}$ & $\tilde{F}_{2, rl}^{(6)}(\tilde{n})$ & $\tilde{S}_{2, rl}^{(6)}(\tilde{n})$ & mean & error    & difference & std-error) \\
\hline
3 &              1              & 0,15625 & 0,15624 & $1 \cdot 10^{-5}$ & $7 \cdot 10^{-6}$ & 0,62 \\
4 &       $\frac{27}{32}$       & 0,29534 & 0,29535 & $1 \cdot 10^{-5}$ & $2 \cdot 10^{-5}$ & 1,13 \\
5 &   $\frac{9\,459}{17\,248}$  & 0,33785 & 0,33784 & $1 \cdot 10^{-5}$ & $1 \cdot 10^{-5}$ & 0,82 \\
6 & $\frac{107\,301}{509\,600}$ & 0,21056 & 0,21056 & $1 \cdot 10^{-5}$ & $3 \cdot 10^{-6}$ & 0,22 \\
\hline
\end{tabular}
\end{center}
\caption{Numerical validation of Eq.~\ref{Eq:ChainedIntegrals}.
The columns $\tilde{F}_{2, rl}^{(6)}(\tilde{n})$ and $\tilde{S}_{2, rl}^{(6)}(\tilde{n})$ refer to analitical values and the columns mean and standard-error came from numeric simulation.
Walks were performed on 300\,000\,000 maps with $N=6$~points each.}
\label{Tab:DistrN6}
\end{table}
}

However, the function $\mbox{max}(x_i, x_k)$ in the lower limits of the integrals in $z_{i, k}$ makes it difficult to solve Eq.~\ref{Eq:ChainedIntegrals}, once we should consider all possible $(\tilde{n}-1)!$ orderings of distances $x_i$.
In the following, we will consider the thermodynamic limit ($N \gg 1$) and make some approximations to solve Eq.~\ref{Eq:ChainedIntegrals}.

For a better visualization, notice that the integrals in $z_{i, k}$ refer to the slashed lines of Fig.~\ref{Fig:Trajectory}.
Observe that exactly $\tilde{n}-4$ slashed lines leave from each site, except the sites $s_1$ and $s_{\tilde{n}-1}$, where $\tilde{n}-3$ slashed lines leave from, due to the additional distance $z_{1, \tilde{n}-1}$ (thick slashed line in Fig.~\ref{Fig:Trajectory}).
To obtain a more regular expression, we can eliminate the integral in $z_{1, \tilde{n}-1}$ in Eq.~\ref{Eq:ChainedIntegrals} (without any harm) and then each variable $z_{i, k}$ appears exactly $\tilde{n}-4$ times.
To justify this elimination notice that, due to the deterministic rule of TW, each distance $x_i$ is the minimum of $N-2$ random variables uniformly distributed in the interval $[0, 1]$.
Therefore, its pdf is given by~\cite{tercariol_2005}: $g(x_i) = (N-2)(1-x_i)^{N-3}$ and its mean and standard deviation are: 
$\overline{x_i} = 1/(N-1) \approx 1/N$ and $\sigma_{x_i} = \sqrt{(N-2)/[N (N-1)^2]} \approx 1/N$, so that, in the limit $N \gg 1$, $x_i$ assumes values close to 0 and the value of the integral $\int_{\mbox{max}(x_1, x_{\tilde{n}-1})}^1 \mbox{d}z_{1, \tilde{n}-1}$ is typically close to 1.

Changing the exponent of $x_{\tilde{n}-1}$ from $N-\tilde{n}$ to $N-\tilde{n}+1$, all the variables $x_i$ are raised to the same power.
The resulting expression is algebraically symmetric with respect to the variables $x_1$, $x_2$, \ldots, $x_{\tilde{n}-1}$, what means that all possible $(\tilde{n}-1)!$ orderings occur with the same probability.
Thus, one can consider the specific ordering $x_1 < x_2 < \cdots < x_{\tilde{n}-1}$ and rewrite Eq.~\ref{Eq:ChainedIntegrals} without using the inconvenient function max():
\begin{eqnarray}
\frac{\tilde{F}_{2, rl}^{(N)}(\tilde{n})}{(\tilde{n}-1)! } = \prod_{i=1}^{\tilde{n}-1} (N-i) \int_0^1 \mbox{d}x_1(1-x_1)^{N-\tilde{n}+1} \nonumber \\
\prod_{i=2}^{\tilde{n}-2} \int_{x_{i-1}}^1 \mbox{d}x_i(1-x_i)^{N-\tilde{n}+i-1} \nonumber \\
\int_{x_{\tilde{n}-2}}^1 \mbox{d}x_{\tilde{n}-1}(1-x_{\tilde{n}-1})^{N-3} \; ,
\label{Eq:IntegralsWithoutMax}
\end{eqnarray}
where we emphasize that the extra factor $(\tilde{n}-1)!$ takes into account all possible orderings of the variables $x_i$.

The exponent of $x_1$ may be changed from $N-\tilde{n}+1$ to $N-\tilde{n}$ aiming the exponents of $x_1$, $x_2$, \ldots, $x_{\tilde{n}-2}$ to be in an arithmetic series.
One then calculates the integrals of Eq.~\ref{Eq:IntegralsWithoutMax} to have at last: 
\begin{eqnarray}
\nonumber
\tilde{F}_{2, rl}^{(N)}(\tilde{n}) & = & \frac{(\tilde{n}-1)! (N-1)(N-2)(N-3)\cdots}{(N-2)(2N-4)(3N-7)(4N-11) \cdots } \nonumber \\
                                   &   & \frac{\cdots(N-\tilde{n}+1)}{\cdots \left\{ (\tilde{n}-1)N-[(\tilde{n}-1)\tilde{n}/2 + 1] \right\} } \nonumber \\
                                   & = & \prod_{k=1}^{\tilde{n}-1} \frac{k(N-k)}{kN - k(k+1)/2 - 1} \nonumber \\
                                   & = & \prod_{j=4}^{\tilde{n}} \frac{N-j+1}{N-j/2-1/(j-1)} \; ,
\label{Eq:DistrTildeNCumul}
\end{eqnarray}
where we have called $j = k + 1$ and the lower limit of the productory was changed from $j=2$ to $j=4$ because the factors for $j=2$ and $j=3$ are physically meaningless, as we shall argue in the Subsec.~\ref{Subsec:AnalogyGeomDistr}.

Finally, the distribution of $\tilde{n}$ is calculated from the one step difference of the upper-tail distribution:
\begin{eqnarray}
\tilde{S}_{2, rl}^{(N)}(\tilde{n}) & = & \tilde{F}_{2, rl}^{(N)}(\tilde{n}) - \tilde{F}_{2, rl}^{(N)}(\tilde{n}+1) \nonumber \\
                                   & = & \left[ 1- \frac{N-\tilde{n}} {N-(\tilde{n}+1)/2 -1/\tilde{n}} \right]  \nonumber \\
                                   &   & \prod_{j=4}^{\tilde{n}} \frac{N-j+1}{N - j/2 - 1/(j-1)} \; .
\label{Eq:DistrTildeN}
\end{eqnarray}
The expression of Eq.~\ref{Eq:DistrTildeNCumul} is similiar to the one obained for $\mu=1$ (using Eqs.~9~and~10 of Ref.~\onlinecite{tercariol_2005} and calling $\tilde{n}=t+2$):
$\tilde{F}_{1, rl}^{(N)}(\tilde{n}) =  [\prod_{j=3}^{\tilde{n}} \frac{N-j+1}{N-j/2}]/(\tilde{n}-1)!$.
The main difference is the presence of the factor $1/(\tilde{n}-1)!$, because, for $\mu=1$, one must to consider only the specific ordering $x_{\tilde{n}-1} < x_{\tilde{n}-2} < \cdots < x_1$.

At this point we are able to understand the major role played by the memory in this partially self avoiding walk. 
For $\mu = 1$, the walker must go to the nearest neighbor. 
The extremal statistics is behind this dynamics. 
But, for instance, forbidding the walker to return to the last visited site, this opens up the possibility to go to the first or second nearest neighbor, which transforms the extremal statistics to the combinatorial statistics. 
Mathematically, this is expressed by the absence of $(\tilde{n} - 1)!$ in Eq.~\ref{Eq:DistrTildeNCumul}.  

\subsection{Analogy to the geometric distribution}
\label{Subsec:AnalogyGeomDistr}

Making an analogy to the geometric distribution, we can write Eq.~\ref{Eq:DistrTildeN} as $\tilde{S}_{2, rl}^{(N)}(\tilde{n}) = \tilde{p}_{2, rl}^{(N)}(\tilde{n}+1) \prod_{j=4}^{\tilde{n}} \tilde{q}_{2, rl}^{(N)}(j)$ where
\begin{eqnarray}
\tilde{q}_{2, rl}^{(N)}(j) = \frac{N-j+1}{N-j/2-1/(j-1)}
\label{Eq:TildeQ}
\end{eqnarray}
is the exploration probability in the $j$th step and $\tilde{p}_{2, rl}^{(N)}(j) = 1 - \tilde{q}_{2, rl}^{(N)}(j)$ is the revisit probability in the $j$th step.
We remark that the expression of Eq.~\ref{Eq:TildeQ} is similar to the one obtained for $\mu=1$ (adapting Eqs.~9~and~10 of Ref.~\onlinecite{tercariol_2005} from their original concept of subsistence probability to the concept of exploration probability handled here): $(j-1)\tilde{q}_{1, rl}^{(N)}(j) =  (N-j+1)/(N-j/2)$. 
The main difference is the extra factor $j-1$, which is a consequence of the restriction $x_{\tilde{n}-1} < x_{\tilde{n}-2} < \cdots < x_1$.
This extra factor explains the abrupt change in the exploratory behavior between $\mu=1$ and $\mu=2$ cases: on one hand, for $\mu=1$ the exploration probability (in the thermodynamic limit) decreases harmonically along the trajectory; on the other hand, for $\mu=2$ the exploration probability tends to 1 when $N \rightarrow \infty$.

Once the memory $\mu=2$ assures the tourist to explore at least $\tilde{n}_{min}=\mu+1=3$~sites, it only makes sense to define exploration probability from the 4th~step. 
In fact, for the first step ($j=1$) Eq.~\ref{Eq:TildeQ} does not have a defined value, for the second step it yields $\tilde{q}_{2, rl}^{(N)}(2) = (N-1)/(N-2) > 1$, which is an absurd, and for the third step $\tilde{q}_{2, rl}^{(N)}(3) = 1$.
To take into account the proper physical content, we previously changed lower limit of the products of Eq.~\ref{Eq:DistrTildeNCumul} from $j=2$ to $j=4$. 
Its interesting to mention that for the step $j=N+1$ (after the tourist explores all $N$ sites), Eq.~\ref{Eq:TildeQ} correctly yields $\tilde{q}_{2, rl}^{(N)}(N+1) = 0$.

Since in the $j$th step there are $j-3$ sites equally probable to be revisited and $\tilde{p}_{2, rl}^{(N)}(j)$ is the probability for the tourist to revisit {\bf any one} of these sites, in the limit $N \gg j \gg 1$ the probability $\tilde{p}_{rl}$ for the tourist to revisit {\bf a specific} site $s_k$ is
\begin{eqnarray}
\tilde{p}_{rl} & = & \frac{1}{j-3} \; \tilde{p}_{2, rl}^{(N)}(j)  =  \frac{1}{j-3} \; \frac{j/2-1-1/(j-1)}{N-j/2-1/(j-1)} \nonumber \\
               & \approx & \frac{1}{2N} \; ,
\label{Eq:TildePrl}
\end{eqnarray}
which is half the probability for he/she to explore a specific new site [namely $\tilde{q}_{rl} = 1/(N-j) \approx 1/N$].

\subsection{Alternative Derivation}

In the following we obtain for $N \gg 1$ simpler expressions for the first passage and exploration probabilities, via an alternative reasoning.
From these probabilities, we obtain closed analytical expressions for $\tilde{F}_{2, rl}^{(N)}(\tilde{n})$.

\subsubsection{First Passage and Exploration Probabilities}

Supose that the tourist has traveled along the trajectory $s_1$, $s_2$, \ldots, $s_{\tilde{n}}$ ($\tilde{n} \ge 3$) without any revisit.
Let us first reobtain the probability $\tilde{p}_{rl}$ for the tourist to revisit a specific site $s_k$ (outside the exclusion window, i.e., $k \le \tilde{n}-2$) in the following step.
To do this, consider the following constraints (see Fig.~\ref{Fig:Trajectory}):
\begin{enumerate}
\item the distance $z_{\tilde{n}, k}$ must be smaller than $x_{\tilde{n}}$.
\item once in the ($k+1$)th step, the tourist came from site $s_k$ to $s_{k+1}$, the distance $z_{\tilde{n}, k}$ is greater than the distance $x_k$.
\end{enumerate}
In brief, $z_{\tilde{n}, k}$ must vary between $x_k$ and $x_{\tilde{n}}$, so that, $0 < x_k < z_{\tilde{n}, k} < x_{\tilde{n}} < 1$.

Once the pdf of each distance $x_i$ is $g(x_i) = (N-2)(1-x_i)^{N-3}$ and $z_{\tilde{n}, k}$ has uniform deviate (by definition of RL model), for $N \gg 1$ the probability $\tilde{p}_{rl}$ is given by: 
$\tilde{p}_{rl}  =  P(x_k < z_{\tilde{n}, k} < x_{\tilde{n}}) = \int_0^1 \mbox{d}x_k (N-2)(1-x_k)^{N-3} \int_{x_k}^1 \mbox{d}x_{\tilde{n}} (N-2)(1-x_{\tilde{n}})^{N-3} \int_{x_k}^{x_{\tilde{n}}} \mbox{d}z_{\tilde{n}, k} \nonumber  =  (N-2)/[(N-1)(2N-3)] \approx 1/(2N)$, which agrees to Eq.~\ref{Eq:TildePrl}. 

For a generic step $j$ there are $j-3$ sites susceptible to be revisited so that the first passage and exploration probabilities for this step are:
$\tilde{p}_{2, rl}^{(N)}(j) = (j-3)/(2N) = 1-\tilde{q}_{2, rl}^{(N)}(j)$, which is an approximation for Eq.~\ref{Eq:TildeQ}, leading to
\begin{eqnarray}
\tilde{F}_{2, rl}^{(N)}(\tilde{n}) & = & \prod_{j=4}^{\tilde{n}} \tilde{q}_{2, rl}^{(N)}(j) = \prod_{j=4}^{\tilde{n}} \left[ 1-\frac{j-3}{2N} \right] \nonumber \\
& = & \frac{\Gamma(2N)}{\Gamma(2N-\tilde{n}+3) (2N)^{\tilde{n}-3}} \; ,
\label{Eq:DistrTildeNGamaCumul}
\end{eqnarray}
which is a closed analytical form for Eq.~\ref{Eq:DistrTildeNCumul}.

\subsubsection{Exponential Form (Cumulative Half Gaussian)}

In the limit $N \gg 1$, the exploration probability may be written as $\tilde{q}_{2, rl}^{(N)}(j) = [ 1-1/(2N) ]^{j-3}$, so that Eq.~\ref{Eq:DistrTildeNGamaCumul} assumes its exponential form
\begin{eqnarray}
\nonumber
\tilde{F}_{2, rl}^{(N)}(\tilde{n}) & = & \prod_{j=4}^{\tilde{n}} \tilde{q}_{2, rl}^{(N)}(j) = \left( 1-\frac{1}{2N} \right)^{\tilde{\omega}} \\ 
                                   & \approx & e^{-\tilde{\omega}/(2N)} = e^{-[(\tilde{n} - 3)^2/(4N) ][1 + 1/(\tilde{n}-3)]} \; ,
\label{Eq:DistrTildeNExp}
\end{eqnarray}
where
\begin{equation}
\tilde{\omega} = \sum_{j=4}^{\tilde{n}} (j-3) = \frac{(\tilde{n} - 2)(\tilde{n} - 3)}{2}
\end{equation}
has a simple physical interpretation.
It is just the number of distances $z_{i, k}$ between non-consecutive sites of trajectory.
Notice that the trajectory of Fig.~\ref{Fig:Trajectory} is topologically equivalent to a ($\tilde{n}-1$)-sided polygon, which has $(\tilde{n}-1)(\tilde{n}-4)/2$ diagonals.
All these diagonals plus the side $s_1 s_{\tilde{n}-1}$ totalize $\tilde{\omega}=(\tilde{n}-2)(\tilde{n}-3)/2$ paths (slashed lines of Fig.~\ref{Fig:Trajectory}), which allow revisit.

For $\tilde{n}-3 \gg 1$, one can disregard $1/(\tilde{n}-3)$ in Eq.~\ref{Eq:DistrTildeNExp}, leading to a half gaussian: $y = \tilde{F}_{2, rl}^{(N)}(\tilde{n}) = e^{-[(\tilde{n}-3)/\sqrt{2N}]^2/2}$, indicating that the scaled variable is $x = (\tilde{n}-\tilde{n}_{min})/\sqrt{2N}$ with $\tilde{n}_{min} = \mu+1 = 3$, leading to the universal curve $y = e^{-x^2/2}$, with $x \ge 0$.
We only have kept $\tilde{n}_{min}$ to compare to a possible generalization of these calculations for the case of short memory $\mu \ll N$.

\section{Distribution of the total number of explored sites}
\label{Sec:DistrN}

Up to this point we were focused on the number $\tilde{n}$ of sites explored before the first revisit.
In the TW with $\mu=1$, the revisit implies the tourist has entered an attractor of period $p=2$~\cite{tercariol_2005}, but with $\mu=2$, the revisit does not implies capture.
In what follows we calculate the probability $p_t$ for the tourist to be trapped during a revisit and then obtain the capture $p_{2, rl}^{(N)}(j)$ and subsistence $q_{2, rl}^{(N)}(j)$ probabilities and the upper-tail distribution $F_{2, rl}^{(N)}(n)$ for the number $n$ of sites visited in the whole walk.

\subsection{Trapping Probability}

Let us recall Fig.~\ref{Fig:Trajectory} and consider that the tourist has traveled along the trajectory $s_1$, $s_2$, \ldots, $s_{\tilde{n}}$ without any revisit.
Assume that in the following step he/she revisits site $s_k$ (outside the memory window, $k \le \tilde{n}-2$).
Due to the deterministic rule, two situations may occur: (i) if $x_k < x_{k-1}$, the tourist moves forward to site $s_{k+1}$ and is trapped by an attractor of period $p=\tilde{n}-k+1$; (ii) if $x_{k-1} < x_k$, the tourist moves backward to site $s_{k-1}$ and escapes from the attractor.
Therefore, the walker trapping or escaping depends on which distance $x_{k-1}$ or $x_k$ is shorter.
The only exception is a revisit to $s_1$, when the tourist is unconditionally trapped, leading to a trajectory with a null transient time ($t=0$) and a cycle of period $p=\tilde{n}$.

Taking into consideration that all $(\tilde{n}-1)!$ possible orderings of the distances $x_1$, $x_2$, \ldots, $x_{\tilde{n}-1}$ are equally probable, one could naively conclude that the trapping probability would be $p_t = P(x_k < x_{k-1}) = 1/2$.
Nonetheless, numerical simulations of this system have refuted this expectation, pointing out that this probability is in fact $p_t = 2/3$.

To understand this result, we first show that the probability $P_v(r)$ for the tourist to revisit a specific site $s_k$ is proportional to the rank $r$ occuped by the associated distance $x_k$ (between sites $s_k$ and $s_{k+1}$) when one reorders the distances $x_1$, $x_2$, \ldots, $x_{\tilde{n}-2}$ decreasingly (so that $x_k$ is the $r$th greatest one).
Secondly, we show that the probability $P_t(r)$ for the tourist to be trapped when revisiting the site $s_k$ is proportional to $r-1$.
Finally, from $P_v(r)$ and $P_t(r)$ we prove that $p_t = 2/3$.

\subsubsection{Order Statistics}

Let us recall some tool about Order Statistics.
Given a sample of $M$ variates $X_1$, $X_2$, \ldots, $X_M$, reorder them so that $X_{(1)} > X_{(2)} > \ldots > X_{(M)}$.
If $X$ has pdf $g(x)$ and cumulative distribution $G(x) = \int_{-\infty}^x dx' g(x')$, then the pdf $h_r(x)$ of $X_{(r)}$ is 
$h_r(x) = M! [G(x)]^{M-r} [1-G(x)]^{r-1} g(x)/[(r-1)!(M-r)!] $, for $r=1, 2, \ldots, M$.

Resuming the TW with $\mu=2$ on the RL model, each distance $x_i$ has pdf given by $g(x)=(N-2)(1-x)^{N-3}$, then its cumulative distribution is $G(x) =\int_0^x \mbox{d}x' \, g(x') = 1-(1-x)^{N-2}$ and the pdf of $x_{(r)}$ is $ h_r(x) = \tilde{n}!  \left[1-(1-x)^{N-2}\right]^{\tilde{n}-r} [(1-x)^{N-2}]^{r-1} (N-2)(1-x)^{N-3}/[(r-1)!(\tilde{n}-r)!]$. 

\subsubsection{Rank-revisit and Rank-trapping Probabilities}

Again, consider that the tourist has traveled along the trajectory $s_1$, $s_2$, \ldots, $s_{\tilde{n}}$ (without any revisit).
Let us calculate the probability $P_v(r)$ for he/she to revisit the site $s_{(r)}=s_k$ (with associated distance $x_{(r)}=x_k$) in the next step.
Once $s_{(r)}$ is the nearest site, the distance $z_{\tilde{n}, (r)}$ has pdf given by $g(x)=(N-2)(1-x)^{N-3}$ and once the tourist came from site $s_{(r)}=s_k$ to $s_{k+1}$ in the ($k+1$)th step, the distance $z_{\tilde{n}, (r)}$ is certainly greater than $x_{(r)}$.
Thus $P_v(r) \propto P(z_{\tilde{n}, (r)} > x_{(r)}) = \tilde{n}!/ [(r-1)!(\tilde{n}-r)!] \int_0^1 \mbox{d}x \left[1-(1-x)^{N-2}\right]^{\tilde{n}-r} \left[(1-x)^{N-2}\right]^{r-1} (N-2)(1-x)^{N-3} \int_x^1 \mbox{d}z (N-2)(1-z)^{N-3}$.
Evaluating the integral in $z$ and calling $y = (1-x)^{N-2}$ the above equation is rewritten as:
$P_v(r) \propto \tilde{n}!/[(r-1)!(\tilde{n}-r)!] \mbox{B}(\tilde{n}-r+1, r+1) = \tilde{n}!/[(r-1)!(\tilde{n}-r)!] (\tilde{n}-r)!r!/(\tilde{n}+1)! = r/(\tilde{n}+1)$.
This expression {\it is not} the probability $P_v$ itself. 
Instead, it only gives the dependence of $P_v$ on $r$.

Normalizing $P_v$ over $1 \le r \le \tilde{n}-2$, one has
\begin{eqnarray}
P_v(r) = \frac{r}{\sum_{k=1}^{\tilde{n}-2} k} = \frac{2r}{(\tilde{n}-1)(\tilde{n}-2)} \; ,
\label{Eq:Pvr}
\end{eqnarray}
where $\tilde{n}-2$ is the number of sites available to revisit (the sites $s_{\tilde{n}}$ and $s_{\tilde{n}-1}$ are forbidden by memory) and the normalization factor $\sum_{j=1}^{\tilde{n}-2} j = (\tilde{n}-2)(\tilde{n}-1)/2$ is simply the sum of all $\tilde{n}-2$ ranks.

The result of Eq.~\ref{Eq:Pvr} does not contradicts Eq.~\ref{Eq:TildePrl}, since Eq.~\ref{Eq:TildePrl} gives an approximated probability for the tourist to revisit a specific site $s_k$, regardless its associated distance $x_k=x_{(r)}$, while Eq.~\ref{Eq:Pvr} gives the conditional probability for the tourist to ``choose'' the $r$-ranked site $s_{(r)}$ during a revisit after exploring $\tilde{n}$ distinct sites.

Once the tourist had revisited site $s_k$ (or equivalently $s_{(r)}$), the probability $P_t(r)$ for he/she to be trapped also depends on the rank $r$.
The trapping condition is that $x_{k-1}$ must be greater than $x_k$.
Since $x_k=x_{(r)}$ is the $r$th greater distance, there are only $r-1$ remaining distances (among $\tilde{n}-3$ ones) greater than $x_k$.
Thus,
\begin{eqnarray}
P_t(r) = \frac{r-1}{\tilde{n}-3} \; .
\label{Eq:Ptr}
\end{eqnarray}

Combining Eqs.~\ref{Eq:Pvr}~and~\ref{Eq:Ptr}, the probability for the tourist to be trapped when visiting a specific site $s_{(r)}$ is:
$P_v(r)  P_t(r) = 2r(r-1)/[(\tilde{n}-1)(\tilde{n}-2)(\tilde{n}-3)]$. 
Thus, the probability for the tourist to be trapped when revisiting any site is $p_t = \sum_{r=1}^{\tilde{n}-2} P_v(r) P_t(r)$. 
Calling $m=\tilde{n}-2$ and evaluating $\sum_{r=1}^m r(r-1) = m(m^2-1)/3$ one finds the trapping probability
\begin{eqnarray}
p_t = 2/3 \; .
\label{Eq:Pt}
\end{eqnarray}

We remark that this result has been obtained without any approximation, and numerical simulations agree to it even for small values of $N$.

\subsection{Capture and Subsistence Probabilities}

Combining the probability $\tilde{p}_{rl}$ for the tourist to revisit a specific site $s_k$ (Eq.~\ref{Eq:TildePrl}) and the trapping probability $p_t$ (Eq.~\ref{Eq:Pt}), one obtains the probability $p_{rl}$ for the tourist to revisit $s_k$ {\bf and} be trapped:
\begin{eqnarray}
p_{rl} = \tilde{p}_{rl} p_t = \frac{1}{2N} \frac{2}{3} = \frac{1}{3N} \; .
\label{Eq:Prl}
\end{eqnarray}
Since in the $j$th step there are $j-3$ sites available to revisit, the capture (i.e., revisiting any site {\bf and} being trapped) and subsistence (i.e., exploring any new site {\bf or} revisiting any site and not being trapped) probabilities in the $j$th step are:
$p_{2, rl}^{(N)}(j) = (j-3)/(3N) = 1 - q_{2, rl}^{(N)}(j)$ and the upper-tail distribution for the number $n$ of sites explored by the tourist in the whole trajectory is
\begin{eqnarray}
F_{2, rl}^{(N)}(n) & = & \prod_{j=4}^n q_{2, rl}^{(N)}(j) \nonumber \\
& = & \frac{\Gamma(3N)}{\Gamma(3N-n+3) (3N)^{n-3}} \; ,
\label{Eq:DistrNGamaCumul}
\end{eqnarray}
which is analogous to Eq.~\ref{Eq:DistrTildeNGamaCumul}.

\subsubsection{Comparison to RM model with $\mu=0$}

The expression of Eq.~\ref{Eq:DistrNGamaCumul} is similar to the one obtained for the RM model with memory $\mu=0$~\cite{tercariol_2005}:
$F_{0, rm}^{(N)}(n)  =  \Gamma(N)/[\Gamma(N-n) N^{n}]$.
This result explains the non-trivial equivalence observed between RL model with $N$ points and memory $\mu=2$ (memory effect) and RM model with $3N$ points and memory $\mu=0$ (effect of distance symmetry break), when one compares the distributions for the total number $n$ of sites explored by the tourist.

Notice that, taking both models with $N$ points each, in RL with $\mu=2$, at each step, the probability for the turist to revisit a specific site and be trapped is $p_{rl} \approx 1/(3N)$; and in RM with $\mu=0$, this probability is $p_{rm} = 1/N$.
Therefore, taking RL with $N$~points and RM with $3N$~points equals these probabilities and justifies the equivalence.

\subsubsection{Exponential Form}

In the limit $N \gg 1$, the subsistence probability is rewritten as $ q_{2, rl}^{(N)}(j) = [1- 1/(3N) ]^{j-3}$ and one obtains the exponential form of Eq.~\ref{Eq:DistrNGamaCumul}, namely $F_{2, rl}^{(N)}(n) = \prod_{j=4}^n q_{2, rl}^{(N)}(j) = [1 - 1/(3N)]^\omega \approx e^{-\omega/(3N)}$, with $\omega = (n-2)(n-3)/2$.

Rather than differentiating $F_{2, rl}^{(N)}(n)$, the distribution $S_{2, rl}^{(N)}(n)$ for the number $n$ of sites explored in the whole trajectory is more precisely obtained by imposing the tourist to explore $n$ distinct sites and then be captured in the next step (i.e., revisit {\bf any} site and be trapped):
\begin{eqnarray}
S_{2, rl}^{(N)}(n) & = & F_{2, rl}^{(N)}(n) \, p_{2, rl}^{(N)}(n+1) \nonumber \\
& = & \frac{n-2}{3N} \, e^{-\frac{(n-2)(n-3)/2}{3N}} \;. 
\label{Eq:DistrNExp}
\end{eqnarray}

\begin{figure}[htb]
\begin{center}
\includegraphics[angle=-90,scale = .35]{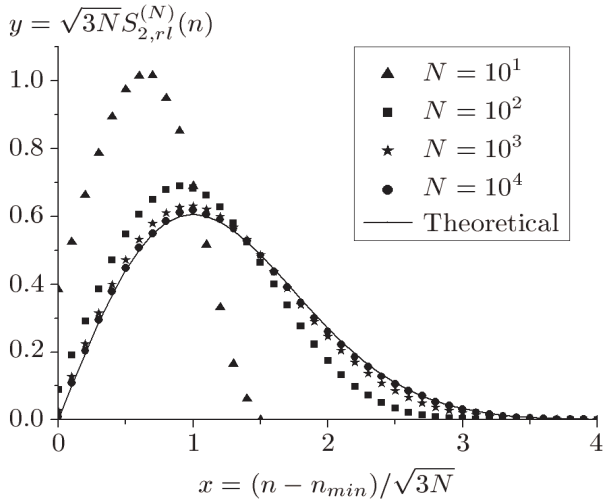}
\caption{Finite size effect for the distributions and convergence to the universal curve: $y = x \, e^{-x^2/2}$, with  $x=(n-n_{min})/\sqrt{3N}$ and $y=\sqrt{3N}S_{2, rl}^{(N)}(n)$.}
\label{Fig:Universal}
\end{center}
\end{figure}

For $n \gg 1$, calling $y = \sqrt{3N} S_{2, rl}^{(N)}(n)$ and $x = (n-n_{min})/\sqrt{3N}$ (with $n_{min}=\mu+1=3$) one obtains the universal plot for this system:
\begin{equation}
y = x \, e^{-x^2/2} \; ,
\label{Eq:Universal}
\end{equation} 
with $x \ge 0$ and $m$th moment $\langle x^m \rangle = 2^{m/2} \Gamma(m/2+1)$, where we see that normalization is assured by $\langle x^0 \rangle = 1$. 
The mean value is $\langle x \rangle = \sqrt{\pi/2}$ and the variance $\langle x^2 \rangle - \langle x \rangle^2 = 2 - \pi/2$.  
Fig.~\ref{Fig:Universal} exhibits a plot of Eq.~\ref{Eq:Universal} and experimental data.
From this figure, or calculating analytically, one obtains that the  mode is unitary. 

\section{Transient and period joint distribution}
\label{Sec:JointDistr}

\begin{figure}[htb]
\begin{center}
\includegraphics[angle=-90, scale=.35]{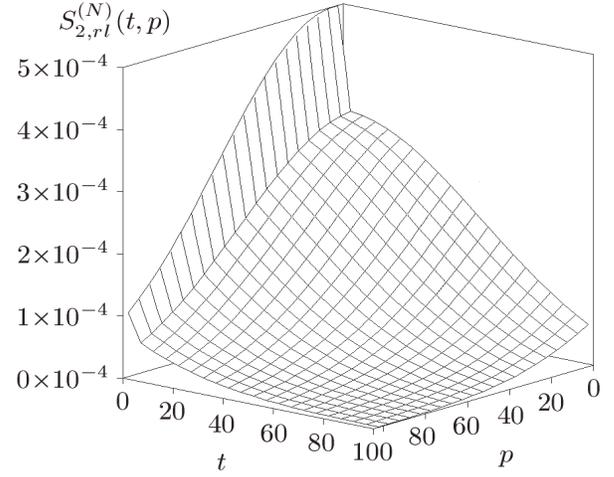}
\caption{Transient and period joint distribution for a map with $N=1000$ points in the RL model with $\mu=2$.}
\label{Fig:JointDistr}
\end{center}
\end{figure}

The transient/period joint distribution $S_{2, rl}^{(N)}(t, p)$ can be obtained similarly to Eq.~\ref{Eq:DistrNExp}, by imposing the tourist to expore $n$ distinct sites and then revist the {\bf specific} site $s_k$ (instead of {\bf any site}) and be trapped, giving rise to a tracjetory with transient $t=k-1$ and period $p=n-k+1$. 
We notice that the relevant variable is $t+p=n$.
Hence, $S_{2, rl}^{(N)}(t, p)$ is obtained multiplying $F_{2, rl}^{(N)}(t+p)$ by $p_{rl}$ (Eq.~\ref{Eq:Prl}) [or by $\tilde{p}_{rl}$ (Eq.~\ref{Eq:TildePrl}) in the case $t=0$, since the tourist is unconditionally captured when revisting the site $s_1$]:
\begin{eqnarray}
S_{2, rl}^{(N)}(t, p) = \frac{1}{(3-\delta_{t, 0})N} \, e^{-\frac{(t+p-2)(t+p-3)/2}{3N}}
\label{Eq:JointDistr}
\end{eqnarray}
where $\delta_{i, j}$ is the Kronecker delta.
Fig.~\ref{Fig:JointDistr} exhibits a plot of Eq.~\ref{Eq:JointDistr} for $N=1000$~points.

\subsection{Transient time marginal distribution}

The transient time distribution is calculated summing Eq.~\ref{Eq:JointDistr} over all possible periods, i.e. $S_{2, rl}^{(N)}(t) = \sum_{p=3}^N S_{2, rl}^{(N)}(t, p)$.
In the limit $N \gg 1$, this summation can be approximated by the integral
\begin{eqnarray*}
S_{2, rl}^{(N)}(t) & = & \int_{5/2}^\infty \mbox{d}p \; S_{2, rl}^{(N)}(t, p) \\
                   & = & \left( 1+\frac{\delta_{t, 0}}{2} \right) \sqrt{\frac{\pi}{6N}} \, \mbox{erfc}\left(\frac{t}{\sqrt{6N}} \right) \; ,
\end{eqnarray*}
where the lower limit $5/2$ is due to a Yates continuity correction (which other than improve the integral approximation, make the analytical form quite simpler) and the upper limit has been extended to infinity to make calculation easier (with no harm, because Eq.~\ref{Eq:JointDistr} yields despicable values for $p>N$) and $\mbox{erfc}(x) = (2/\sqrt{\pi}) \int_z^\infty \mbox{d}x \, e^{-x^2}$ is the complementary error function.

\subsection{Cycle period marginal distribution}

Similarly, the period distribution is
\begin{eqnarray*}
S_{2, rl}^{(N)}(p) & = &\sum_{t=0}^{N-3} S_{2, rl}^{(N)}(t, p) = \int_{-1}^\infty \mbox{d}t \; S_{2, rl}^{(N)}(t, p) \\
                   & = &\sqrt{\frac{\pi}{6N}} \, \mbox{erfc}\left(\frac{p-7/2}{\sqrt{6N}} \right) \approx \frac{e^{-p^2/(6N)}}{p} \; ,
\end{eqnarray*}
where the lower limit $-1$ is due to both Yates continuity correction and a compensation for the half extra degree in $t=0$.
The mean period value is $\overline{p} = \sqrt{3 \pi N/8}$ and standard deviation is $\sigma_p = \sqrt{(2 - 3 \pi/8) N}$. 
For $p \ll \sqrt{6N}$, the decay follows a power law $S_{2, rl}^{(N)}(p) \propto p^{-1}$.

\section{Conclusion}
\label{Sec:Conclusion}

In this paper, we have analytically obtained the statistical distributions for the deterministic tourist walk with memory $\mu=2$ on the random link model.

The distribution for the number of sites explored before the first passage has been compared to the one previously obtained for the case $\mu=1$, elucidating the mechanism that strongly increases the tourist's exploratory behavior. 
On one hand, for $\mu=1$ the distances travelled at each step must obey $x_1>x_2>\ldots$, leading to a localized exploration.
In the thermodynamic limit, the mean number of explored sites is then $\overline{n}= e = 2.71828 \ldots$ and the exploration probability decreases harmonically along the trajectory.
This dynamics is due to the underlying extremal statistics.
On the other hand, for $\mu=2$ the distances $x_1, x_2, \ldots$ are unconstrained, leading to an extended exploration: $\overline{n}$ is proportional to $N^{1/2}$ and the exploration probability tends to 1, when $N \rightarrow \infty$.
This dynamics  is due to the underlying combinatorial statistics.
The factor $(\tilde{n}-1)!$ in Eq.~\ref{Eq:IntegralsWithoutMax} represents the change from the extremal statistics to the combinatorial one, which makes the $\delta_{p,2}$ distribution of $\mu = 1$ to broaden to a wide ($1/p$) distribution for $\mu \ge 2$.

Throught the trapping probability $p_t=2/3$ (which value is counterintuitive), we obtained the capture and subsistence probabilities and a closed form to the complementary cumulative distribution for the number of sites explored in the whole trajectory.
This distribution is analogous to the one obtained for the random map model with $\mu=0$.
This result explains the equivalente between these mean field models (RL with $N$ points and memory $\mu=2$; and RM with $3N$ points and memory $\mu=0$).

For a large number of sites ($N \gg 1$) in the random medium, the distribution $S^{(N)}_{2, rl}(n)$ of having $n$ distinct sites visited by the tourist with memory $\mu = 2$ in the random link model is universal $y = x \, e^{-x^2/2}$ with $y = \sqrt{3N} S^{(N)}_{2,rl}(n)$ and $x = (n - 3)/\sqrt{3N}$. 

The transient time $t$ and cycle period $p$ joint distribution $S^{(N)}_{2, rl}(t, p) = e^{[(t+p-3)^2/(3N)]/2}/[N(3 - \delta_{t, 0})]$ has been obtained noticing that the revelant variable is approximatively given by $t+p=n$. 
The marginal distributions are also universal. 
For the transient time one has: $y = [1 + \delta(x)/2]\mbox{erfc}(x)$ with $y = \sqrt{6N/\pi} S^{(N)}_{2,rl}(t)$ and $x = t/\sqrt{6N}$ and for the period distribution: $y = \mbox{erfc(x)}$, with $y = \sqrt{6N/\pi} S^{(N)}_{2,rl}(p)$ and $x = (p - 7/2)/\sqrt{6N}$. 
We have shown that the discrepance in the null transient time distribution ($t=0$), when compared to the subsequent ones ($t>0$), is due to the higher capture probability the starting site $s_1$ has [namely, $\tilde{p}_{rl}=1/(2N)$] when compared to the others else [$p_{rl}=1/(3N)$].
We also have shown that the period distribution decays according a power law $S^{(N)}_{2,rl}(p) \propto p^{-1}$.

Future studies concern the consideration of higher memory values in the random link model and the understanding of the connection with the random map model. 
As the memory increases, we expect a transition from the closed periods to non-closed ones (chaotic phase). 
We are interested in understanding the role of finite dimensionality of the system.

\section*{Acknowledgements}

The authors thank R. S. Gonzalez for fruitful discussions.
ASM acknowledges the Brazilian agencies CNPq (303990/2007-4 and 476862/2007-8) for support.


\end{document}